\documentclass[aps,prb,twocolumn,superscriptaddress,showpacs]{revtex4}
\usepackage[dvips]{graphicx,color}
\begin{document}
\title{Superconductivity induced by U-doping in the SmFeAsO system}
%%%%%%%%%%%%%%%%%%%%%%%%%%%%%%%%%%%
\author{Bo Huang}
\affiliation{Key Laboratory of Radiation Physics and Technology (Sichuan University),
Ministry of Education; Institute of Nuclear Science and Technology, Sichuan University\\
Chengdu 610064, P. R. China.\\}

%%---------------------------
\author{Jijun Yang}
\affiliation{Key Laboratory of Radiation Physics and Technology (Sichuan University),
Ministry of Education; Institute of Nuclear Science and Technology, Sichuan University\\
Chengdu 610064, P. R. China.\\}

\author{Jun Tang}\email{tangjun@scu.edu.cn}
\affiliation{Key Laboratory of Radiation Physics and Technology (Sichuan University),
Ministry of Education; Institute of Nuclear Science and Technology, Sichuan University\\
Chengdu 610064, P. R. China.\\}

\author{Jiali Liao}
\affiliation{Key Laboratory of Radiation Physics and Technology (Sichuan University),
Ministry of Education; Institute of Nuclear Science and Technology, Sichuan University\\
Chengdu 610064, P. R. China.\\}

\author{Yuanyou Yang}
\affiliation{Key Laboratory of Radiation Physics and Technology (Sichuan University),
Ministry of Education; Institute of Nuclear Science and Technology, Sichuan University\\
Chengdu 610064, P. R. China.\\}

\author{Ning Liu}
\affiliation{Key Laboratory of Radiation Physics and Technology (Sichuan University),
Ministry of Education; Institute of Nuclear Science and Technology, Sichuan University\\
Chengdu 610064, P. R. China.\\}

%%---------------------------
\author{Gang Mu}\email{mugang@mail.sim.ac.cn}
\affiliation{State Key Laboratory of Functional Materials for
Informatics, Shanghai Institute of Microsystem and Information
Technology, Chinese Academy of Sciences\\
865 Changning Road, Shanghai 200050, China\\
}

%%---------------------------
\author{Tao Hu}
\affiliation{State Key Laboratory of Functional Materials for
Informatics, Shanghai Institute of Microsystem and Information
Technology, Chinese Academy of Sciences\\
865 Changning Road, Shanghai 200050, China\\
}

%%--------------------------
\author{Xiaoping Shen}
\affiliation{State Key Laboratory of Surface Physics, Department of Physics, and Advanced Materials Laboratory,
Fudan University, Shanghai 200433, China\\
}

%%---------------------------
\author{Donglai Feng}
\affiliation{State Key Laboratory of Surface Physics, Department of Physics, and Advanced Materials Laboratory,
Fudan University, Shanghai 200433, China\\
}

\date{\today}
%%%%%%%%%%%%%%%%%%%%%%%%%%%%%%%%%%%%%%
\begin{abstract}
Through partial substitution of Sm by U in SmFeAsO, a different member of the family of 
iron-based superconductors was successfully synthesized. X-ray
diffraction measurements show that the lattice constants along
$a$-axis and $c$-axis are both squeezed through U doping, indicating
a successful substitution of U at the Sm site. The parent compound shows
a strong resistivity anomaly near 150 K, associated with
spin-density-wave (SDW) instability. U doping suppresses this
instability and leads to a transition to the superconducting state
at temperature up to 49 K. Magnetic measurements confirm the bulk
superconductivity in this system. For the sample with a doping level
$x$ = 0.2, the external magnetic field suppresses the onset temperature
very slowly, indicating a rather high upper critical field. In
addition, the Hall effect measurements show that U clearly dopes
electrons into the material.

\end{abstract}

%\keywords{Iron arsenide, oxypnictide,uranium, unconventional superconductor}
            %Use show keys class option if keyword
            %Display desired
\maketitle
%%%%%%%%%%%%%%%%%%%
%%=================================================
%% Introduction%  Background & History
%% problems and others
%%%----------------------------
%% purpose of the present study
%%%=== History

The discovery of iron-based superconductors spawned a new research
upsurge of superconductors~\cite{Kamihara}. These compounds became
ideal superconductors due to their amazing properties different from
conventional BCS superconductors and the cuprates. The huge family
of iron-based superconductors can be classified as 1111 type, 122
type, 111 type, 11 type, 21311 type, and so on, according to their
structure features. Among these categories, the 1111-type compounds
are found to has a tetragonal ZrCuSiAs-type crystal structure and
the highest critical transition temperature ($T_c$). Different
element substitutions have been successfully performed by
researchers to induce superconductivity in this system, e.g.,
O$\rightarrow$F~\cite{Kamihara,CeFeAsO_F,PrFeAsO_F,NdFeAsO_F,SmFeAsO_F,GdFeAsO_F},
Ln (Ln = rare-earth
elements)$\rightarrow$Th~\cite{Gd_ThFeAsO,Tb_ThFeAsO,Nd_ThFeAsO}/Sr~\cite{La_SrFeAsO,Nd_SrFeAsO,Pr_SrFeAsO},
Fe$\rightarrow$Co~\cite{LaFe_CoAsO,SmFe_CoAsO}/Ni~\cite{LaFe_NiAsO}/
Ir~\cite{SmFe_IrAsO,LaFe_IrAsO}, As$\rightarrow$P~\cite{LaFeAs_PO}.
Generally speaking, the effect of element substitution is to adjust
the charge carrier concentration and the structural parameters.
These two factors seem to be crucial to enhance $T_c$ in iron-based
superconductors~\cite{Structural Parameter}. The indirect
substitution (e.g. O$\rightarrow$F) is found to have more advantages
in achieving a higher $T_c$ than the direct substitution (e.g.
Fe$\rightarrow$Co). However, it is difficult to control the amount
of fluorine in the system, especially in the process of
single-crystal growth~\cite{single-crystal growth}. Similar problem
also remains for other indirectly doped 1111-systems, $i.e.$, the
alkaline-earth elements (Sr, Ca) substitution at the Ln sites
(Ln=La, Pr, Nd)~\cite{La_SrFeAsO,Nd_SrFeAsO,Pr_SrFeAsO}. Doping on
the site of Ln (Ln=Gd, Tb, Nd) by the 5$f$ element thorium (Th) was
found to be another effective indirect substitution, which can also
induce superconductivity with $T_c$ above 50 K in
Ln$_{1-x}$Th$_x$FeAsO~\cite{Gd_ThFeAsO,Tb_ThFeAsO,Nd_ThFeAsO}.
However, there is still a large space to improve the volume fraction
of magnetic shielding in Th-doped systems.

Obviously, it is requisite to explore alternative ways to introduce
high-$T_c$ superconductivity in a easily-controlled way and with a
high volume fraction. It has been found that the FeAs$_4$-lattices
in SmFeAsO can form a regular tetrahedron naturally, which is
profitable to obtain the high $T_c$~\cite{Structural Parameter}. In
this sense, another 5$f$ element uranium (U) is expected to be a
better dopant on Sm site, because the radius of U$^{4+}$ is very
close to that of Sm$^{3+}$ so that to minimize the lattice mismatch.
In this Rapid Communication, we report the successful synthesis and
characterization of another 5$f$ element£¬ the uranium-doped
superconducting material Sm$_{1-x}$U$_x$FeAsO with $T_c$ as high as
49 K. Bulk superconductivity is confirmed by resistivity and
magnetization measurements. A clear evolution of the charge carriers
is evidenced by the Hall effect data. Our results suggest that there
many opportunities to find different superconductors in this field.

%%Obviously, there stands a good chance that using other actinide elements to dope in
%%the rare-earth sites will cause superconductivity with high $T_c$ as well.
%%But actinide elements are all artificially synthesized except Ac, Th, Pa and U, and
%%among these four elements there are only Th and U have a long half-time. Except for
%%this reason above, U is 5$f$ element, hence it's interesting to explore
%%whether it would induce superconductivity
%%when 4$f$ element Sm are partially replaced by 5$f$ element U. Therefore in this work
%%we use U$^{4+}$ which has a more similar ionic radii to Sm$^{3+}$
%%than Th$^{4+}$ to dope in SmFeAsO, and observed a superconductivity with $T_c$
%%as high as 49 K. X-ray diffraction (XRD) pattern, dc magnetization, resisitivity,
%%upper critical field and Hall effect have been determined and discussed.

The polycrystalline samples of Sm$_{1-x}$U$_x$FeAsO were synthesized
by solid state reaction in an evacuated quartz tube. The starting
materials are SmAs, U$_3$O$_8$, FeAs, Fe and Fe$_2$O$_3$, with the
purity of higher than 99.95\%. SmAs and FeAs were presynthesized by
reacting Sm tapes and Fe powders with As grains at high temperature.
U$_3$O$_8$ was acquired from the decomposition of UO$_2$(NO$_3$)$_2$
under 800 $^\circ$C. The raw materials were mixed well according to
the stoichiometry, and then pressed into pellets in an argon-filled
glove box (both H$_2$O and O$_2$ are limited below 0.1 ppm). The
pressed pellets were sealed in an evacuated quartz tube, and slowly
heated to 1180 $^\circ$C, holding for 90 h. Finally the samples were
furnace cooled to room temperature.

%%%===***********************************==

Powder x-ray diffraction (XRD) measurements was performed at room
temperature with DX-2500 X-ray diffractometer using Cu K$\alpha$
radiation from 20$^\circ$ to 80$^\circ$ with a step of 0.03$^\circ$.
The dc magnetization was measured with a superconducting quantum
interference device (Quantum Design, MPMS-7T). The resistivity and
Hall effect measurements were performed using a physical property
measurement system (Quantum Design,PPMS-9T) with magnetic fields up
to 9 T. The electrical resistivity was measured by a standard
four-probe technique using silver paste electrodes.

%%%=====================================
%%%===***********************************==
\begin{figure}[ht]
\includegraphics*[width=1.1\linewidth]{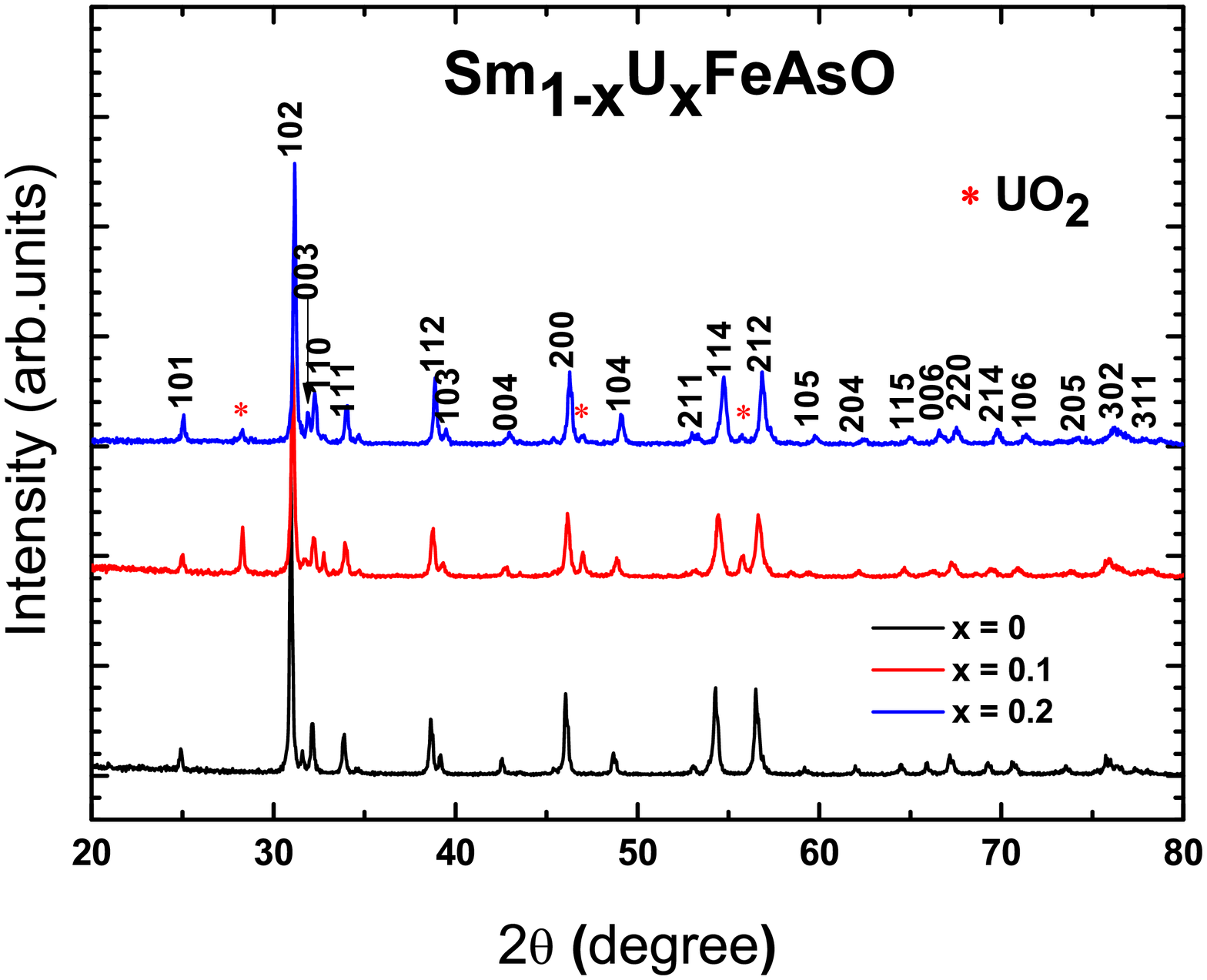}
\caption{(Color online) X-ray diffraction patterns for
Sm$_{1-x}$U$_x$FeAsO samples with doping level $x$ = 0, 0.1 and
0.2. Small amount of impurity phases have been detected on the
doped samples, which are denoted by red asterisks. }\label{St}
\end{figure}
%%%---------------------------------------

Figure 1 shows the powder XRD pattern of Sm$_{1-x}$U$_x$FeAsO ($x$ =
0, 0.1, 0.2) samples. The main diffraction peaks of the three
samples can all be well indexed based on the tetragonal
ZrCuSiAs-type structure with the space group P4/nmm. The refined
lattice parameters are $a$=3.9396(7) {\AA} and $c$=8.4921(4) {\AA}
for the parent compound, basically consistent with the previously
reported values~\cite{SmFeAsO_F}. The tiny peaks marked by the
asterisks are contributed by the secondary UO$_{2}$ phase, which is
the main impurity phase in the U-doped samples. The presence of
impurity phase also indicates that the actual doping level is less
than nominal one. We note here that all the doping values in this
paper are nominal ones. The refined lattice constants for $x$=0.1
and 0.2 are $a$ = 3.9315(7) {\AA}, $c$ = 8.4654(2) {\AA} and $a$ =
3.9214(2) {\AA}, $c$ = 8.4165(7) {\AA}, respectively. We can see
that, by substituting U into Sm sites, both the lattice constants
along $a$ and $c$ axis shrink. This tendency is similar to the case
of doping F into O site in the Ln$_2$O$_2$
layers~\cite{Kamihara,CeFeAsO_F,PrFeAsO_F,lattice1}, but different
from the case of substituting Ln by Th
atoms~\cite{Gd_ThFeAsO,Tb_ThFeAsO,Nd_ThFeAsO}, where Th substitution
expands the crystal lattice along $a$ axis and squeezes it along $c$
axis. This phenomenon is most likely due to the different relative
radii of the doped ions. We emphasize that the substantial and
systematical change in lattice constants indicates that uranium is
incorporated in the lattice successfully.

%%%%%%%%%%%%%%%%%%%%%%
\begin{figure}[ht]
\includegraphics*[width=0.8\linewidth]{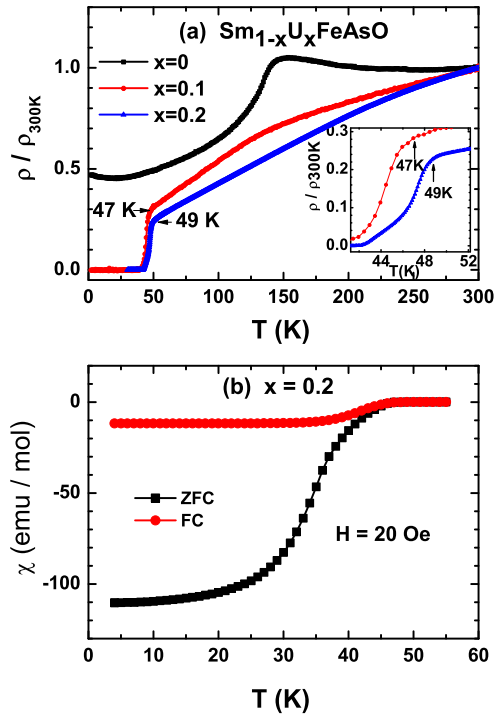}
\caption{(Color online) (a) Temperature dependence of resistivity
for three Sm$_{1-x}$U$_x$FeAsO samples in temperature up to 300 K.
(b) Temperature dependence of magnetic susceptibility of one U-doped
sample with $x$ = 0.2 in ZFC and FC situations under applied field
of 20 Oe. }\label{St}
\end{figure}
%%%%%%%%%%%%%%%%%%%%
%%\subsection{C.Resistivity}
The resistivity of Sm$_{1-x}$U$_x$FeAsO is shown in Fig. 2(a) as a
function of temperature. The resistivity of the undoped sample shows
an anomaly at about 150 K, which is a quite common feature in the
1111-type parent compounds associated with structural phase
transition and antiferromagnetic spin-density-wave (SDW)
instability~\cite{SDW1,SDW2}. Such an anomalous feature is
suppressed by U doping, accompanying the emergence of
superconductivity in low temperatures. From the data we can see the
onset transition temperatures for the samples with $x$ = 0.1 and 0.2
are 47 K and 49 K, respectively, as clearly revealed in the inset of
Fig. 2(a). For $x$ = 0.1, the resistivity in the normal state shows
a linear dependence with temperature below about 120 K, while a
clear negative curvature is observed in the wide temperature range.
This behavior has been reported in other electron-doped 1111
systems~\cite{Gd_ThFeAsO,SmFeAsO anomaly,Nd-1111crystal}, which may
reflect the presence of unconventional scattering mechanism in this
system. With the increase of doping content, the negative-curvature
feature is weakened, as shown in the data with $x$ = 0.2. Such an
interesting evolution needs more investigations in the future.

%%%****************************
\begin{figure}[ht]
\includegraphics*[width=0.9\linewidth]{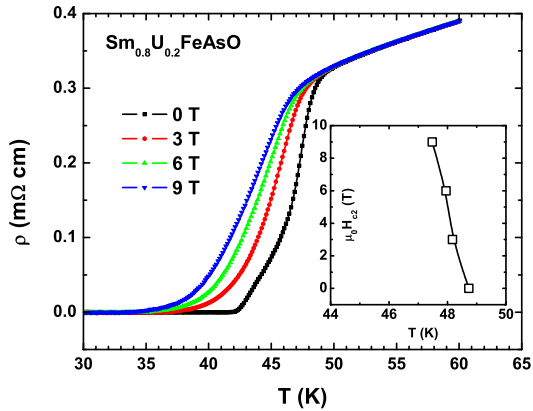}
\caption{(Color online) Temperature dependence of resistivity for
the sample with $x$ = 0.2 under different magnetic fields. The
inset shows the phase diagram derived from the resistive transition
curves taking a criterion of 95\% $\rho_n$.}\label{St}
\end{figure}
%%%%%%%%%%%%%%

The U-doped samples were checked by magnetic susceptibility
measurements. Here we show the data for $x$ = 0.2 in Fig. 2(b) as an
example. The measurements were carried out under a magnetic field of
20 Oe in zero-field-cooled and field-cooled processes. One can see
clear diamagnetic signals in the low temperature region. The onset
transition temperature is determined to be about 46 K for this
sample, which corresponds well to the middle point of the resistive
transition. The magnitude of the dc susceptibility suggests the bulk
superconductivity in the present sample. We note here that the
volume fraction of magnetic shielding in the present system is
clearly improved compared with the Th-doped
systems~\cite{Gd_ThFeAsO,Tb_ThFeAsO,Nd_ThFeAsO}.

%%\subsection{D.Upper critical field}
To obtain the information of upper critical field, the resistivity
data near the superconducting transition were collected under
different magnetic fields. As shown in Fig. 3, the transitions for
the sample with $x$ = 0.2 are suppressed gradually by the field. It
has been pointed out that the onset transition point mainly reflects
the upper critical field in the configuration of
H$\|$ab-plane~\cite{critical field 1}. Taking a criterion of 95\%
$\rho_n$, the onset transition temperatures under different fields
are determined and plotted in the inset of Fig. 3. From these data,
we can get the slope of $H_{c2}$(T) near $T_c$,
$(dH_{c2}/dT)_{T=T_c}\approx$ -7.4 T/K. This value is comparable to
other 1111 system with similar $T_c$~\cite{critical field
2,critical field 3}. Then the simple Werthamer-Helfand-Hohenberg
(WHH) formula~\cite{WHH}, $H_{c2}(0)=-0.693\times
T_c(dH_{c2}/dT)|_{T=T_c}$, is adopted to estimate the upper critical
field to be 250 T.
%%%===***********************************==

%%%%%%%%%%%%%%%%%%%%%%%%%
\begin{figure}[ht]
\includegraphics*[width=0.75\linewidth]{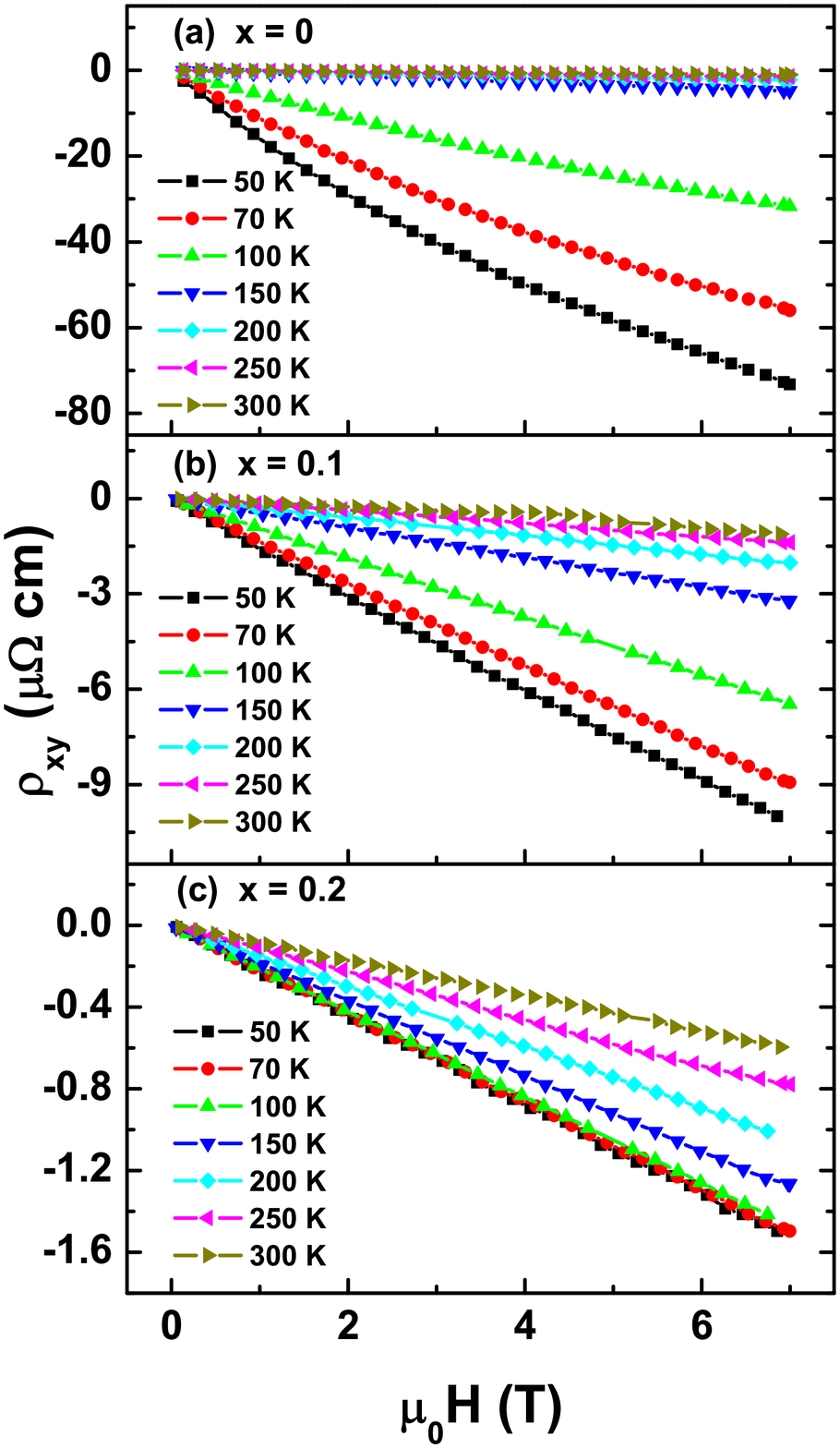}
\caption{(Color online) Field dependence of Hall resistivity
$\rho_{xy}$ in the normal state for three Sm$_{1-x}$U$_x$FeAsO
samples with $x$ = 0 (a), 0.1 (b), and 0.2 (c). }\label{St}
\end{figure}
%%%%%%%%%%%%%%%%%%%%%%%%%%%%%%%%%%%%%%%
%%%%Hall effect
%%%****************************
\begin{figure}[ht]
\includegraphics*[width=0.95\linewidth]{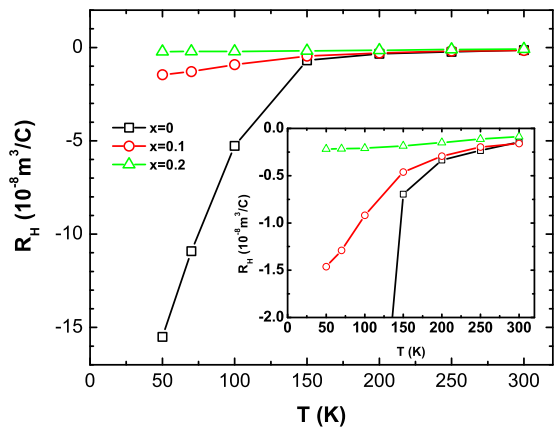}
\caption{(Color online) Temperature dependence of Hall coefficient
$R_H$ for the samples Sm$_{1-x}$U$_x$FeAsO with $x$=0, 0.1 and 0.2.
The inset shows the enlarged view of the same data. }\label{St}
\end{figure}
%%%%%%%%%%%%%%

In order to get more information about the conducting carriers in
the samples, we also carried out the Hall-effect measurements.
Figure 4 shows the magnetic field dependence of Hall resistivity
$\rho_{xy}$ at different temperatures for the samples with $x$ = 0,
0.1, and 0.2, respectively. The value of $\rho_{xy}$ was taken as
$\rho_{xy}$ = [$\rho$(+H) - $\rho$(-H)]/2 at each point to eliminate
the effect of the misaligned Hall electrodes. For the samples with
$x$ = 0.1 and 0.2, almost all the curves show a good linearity
versus the magnetic field and $\rho_{xy}$ is negative at all
temperatures above the critical temperature, indicating domination
of electron-like charge carriers in the normal-state conduction. For
the parent sample, however, the linear behavior can only be observed
in the temperatures above 150 K, suggesting a drastic change of the
conducting conditions during the structural and antiferromagnetic
transitions at 150 K.

Temperature dependence of Hall coefficient $R_H$ is shown in Fig. 5.
It is clear that the absolute value of $R_H$ decreases with
increasing $x$, indicating that U doping leads to an increase of the
electron concentration. For the sample $x$ = 0, the Hall coefficient
shows a pronounced change at about 150 K, which can also be
originated from the structural and antiferromagnetic transitions. We
note that $R_H$ of the superconducting samples also shows a sizable
temperature dependent behavior, as enlarged in the inset of Fig. 5.
This may reveal a trace of multi-band effect in the present system.

In summary, we found that U doping is effective in inducing
superconductivity in SmFeAsO system. The highest critical transition
temperature of about 49 K was confirmed by resistivity and magnetic
susceptibility measurements. X-ray diffraction patterns indicate
that the obtained materials have formed the ZrCuSiAs-type crystal
structure. The systematic evolution of the lattice constants with
doping demonstrated that the U ions have been successfully doped
into the sites of Sm. The upper critical field is estimated for the
sample with $x$ = 0.2. Hall effect measurements show a clear
electron-doping process with the increasing U concentration. Our
results demonstrate a new and better approach to introduce
superconductivity in the 1111 system, where the doping content is
easier to control and the volume fraction of magnetic shielding is
improved. The present system may be suitable for the single-crystal
growth and in-depth research in the next step.
%%%********************************************

%%=================================================

%%%%%%%%%%%%%%%%%%%
This work was supported by National Natural Science Foundation of
China (No. 91226108, 11274234 and 11204338), the International
Thermonuclear Experimental Reactor (ITER) Program Special (Grant No.
2011GB108005), the Program for New Century Excellent Talents in
University (No. NCET-10-0571), the Knowledge Innovation Project of
Chinese Academy of Sciences (KJCX2-EW-W11), the National Fund of China
for Fostering Talents in Basic Science (J1210004), and State Key Laboratory
of Surface Physics, Fudan University (No.KS2011-02).

\end{document}